# Photophysics of Deep Blue Acridane- and Benzonitrile-Based Emitter Employing Thermally Activated Delayed Fluorescence

*Nikita A. Drigo,[†,⊥] Liudmila G. Kudriashova,[‡,⊥] Sebastian Weissenseel,[‡] Andreas Sperlich,[‡] Aron Joel Huckaba,[†] Mohammad Khaja Nazeeruddin,\*,[†] and Vladimir Dyakonov\*[‡,§]*

[†] Group for Molecular Engineering of Functional Materials, École polytechnique fédérale de Lausanne (EPFL), CH-1951 Sion, Switzerland

[‡] Experimental Physics 6, Julius-Maximilian University of Würzburg, 97074 Würzburg, Germany

[§] Bavarian Center for Applied Energy Research (ZAE Bayern), 97074 Würzburg, Germany

[⊥] Denotes an equal contribution.

AUTHOR INFORMATION

**Corresponding Authors**

\* E-mail: mdkhaja.nazeeruddin@epfl.ch (M.K.N).

\* E-mail: dyakonov@physik.uni-wuerzburg.de (V.D.).

ABSTRACT. We designed and synthesized a new organic light-emitting diode (OLED) emitter, SBABz4, containing spiro-biacridine donor (D) in the core surrounded by two benzonitrile acceptors (A). The dual A-DxD-A structure is shown to provide pure-blue emission in relation to its single A-D counterpart. Time-resolved photoluminescence (TRPL) recorded in the broad dynamic range from solutions and solid films revealed three emission components: prompt fluorescence, phosphorescence, and efficient thermally-activated delayed fluorescence (TADF). The latter is independently proven by temperature-dependent TRPL and oxygen-quenching PL experiment. From the PL lifetimes and quantum yield, we estimated maximum external quantum efficiency of 7.1% in SBABz4-based OLEDs, and demonstrated 6.8% in a working device.

**TOC GRAPHICS**

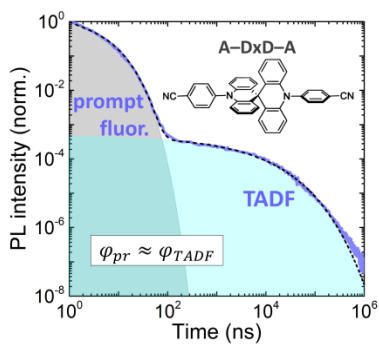

**KEYWORDS.** Deep blue fluorescence, organic light-emitting diodes, spiro compounds, thermally activated delayed fluorescence, time-resolved spectroscopy.

Thermally activated delayed fluorescence (TADF) changed the field of organic optoelectronics since the first reports by Adachi and co-authors in 2011.[1,2] Internal electroluminescence quantum efficiencies close to 100% and outstanding external quantum efficiencies (EQE) above 30% were achieved in the number of organic light-emitting diodes (OLEDs) employing metal-free TADF emitters.[3–8] This tremendous research progress was caused by high application potential of the new OLED generation, as well as by fascinating TADF photophysics.

Electrical injection of carriers produces singlet and triplet excitons with 1:3 branching ratio, reducing the emission probability to 25% of the entire excited population in organic semiconductors. Therefore, maximum internal electroluminescence efficiency of conventional OLED emitters does not exceed 25%, imposing a theoretical limit of 5–7.5% on EQE of the devices with 20–30% light outcoupling. Along with the triplet-triplet annihilation[9,10] and heavy-metal-complex phosphorescence,[11–14] TADF was shown to take advantage of dark triplet states, subsequently improving internal electroluminescence efficiency. TADF opened the way to fully overcome the efficiency-limiting spin statistics by thermally activated triplet-to-singlet up-conversion.

The efficiency of the triplet up-conversion, which occurs via reverse intersystem crossing (RISC), increases with decreasing singlet-triplet energy splitting, $\Delta E_{ST}$.[15] $\Delta E_{ST}$ is commonly rendered low by a large dihedral angle between donor (D) and acceptor (A) moieties and consequently small HOMO–LUMO spatial overlap. This conformational requirement significantly limits the variety of possible donor-acceptor pairs for intramolecular TADF. On the other hand, energies of the HOMO and LUMO, together with the spatial overlap of the orbitals, define the optical properties essential for applications, emission color in particular. Molecular design of pure blue emitters is especially challenging since they should combine wide optical gap, narrow luminescence spectrum, and sufficient radiative efficiency. Therefore, despite the vast demand of organic optoelectronics, deep blue TADF emitters remain scarce.[5,7,16–18]

The majority of intramolecular TADF emitters currently presented in literature have a D-A-D structure with one or several donor moieties attached to a single acceptor in the core. Alternatively, molecules with dual emissive cores were shown to be superior in performance to their single-core analogs.[19,20] Moreover, several spiro-linked donors were developed to provide rigid stick-like structure and partial self-orientation of the molecules.[5,6] This approach enhanced light outcoupling, resulting in an efficiency breakthrough in the blue TADF-based OLEDs. However, photophysics of TADF emitters with a double donor in the core, as well as the role of spiro-linkage is yet to be revealed.

Herein we present a novel compact deep-blue TADF emitter with a spiro-biacridine double donor core surrounded by two benzonitrile acceptors. We determine the correlation between structure and luminescence properties of this dual A-DxD-A compound in relation to its single A-D counterpart. The spiro-bi-donor shifts the emission to the deep blue region. The blue emitter combines phosphorescence at low temperature with TADF at room temperature. We show that the emitter is applicable for OLEDs and demonstrates $EQE_{max} = 6.8\%$ in a test device. We also compared photophysical properties of neat emitters with those diluted in a (bis[2-(diphenylphosphino)phenyl]ether oxide) (DPEPO) matrix to study the luminescence quenching effects and the influence of molecular environment.

Figure 1a shows the molecular structures of 4,4'-(10$H$,10'$H$-9,9'-spirobi[acridine]-10,10'-diyl)dibenzonitrile (SBABz4) and its monomer counterpart, 4-(9,9-dimethylacridin-10(9$H$)-yl)benzonitrile (DMABz4). The deep HOMO of SBA or DMA donor was combined with shallow LUMO of benzonitrile acceptor to provide a wide optical gap sufficient for emission in the blue-green region.[5,21,22] Mutual arrangement of H-atoms at the *peri*-position of the acridine and at *o*-position of the acceptor's phenyl ring impedes D-A rotation, resulting in a semi-frozen conformation with a large dihedral angle between the acridine and benzonitrile planes.[21] This nearly orthogonal D-A orientation is expected to break the HOMO-LUMO spatial overlap, thus decreasing the $\Delta E_{ST}$ and giving rise to TADF.

We prepared the SBA core using a modified procedure by Ooishi et al.[23] The optimized synthesis was carried out on a milligram-to-gram-scale with shortened reaction times and improved total yield. DMA core was synthesized according to the published protocol.[24] Buchwald-Hartwig amination with 4-bromobenzonitrile concluded the preparation of the desired emitters. All synthetic details, as well as thermal and electrochemical characterization of the obtained compounds are provided in the Supporting Information (SI).

Figure 1b displays normalized photoluminescence (PL) and electroluminescence (EL) spectra of SBABz4, recorded in a neat film, a doped 5wt% SBABz4:DPEPO matrix,[25] dilute dichloromethane (DCM) solution, and SBABz4-based OLED. SBABz4 has pure blue emission peaking at 435 nm for the neat film. EL from the SBABz4-based device is in excellent agreement with the PL from the corresponding doped matrix. The structureless luminescence spectra with significant positive solvatochromism (Figure S4, Table S1) indicate the emission from CT state.[26]

Figure 1b compares PL spectra of DMABz4 and SBABz4 dilute solutions in DCM. Whereas SBABz4 shows blue emission ($\lambda_{max} = 470\ nm$), emission from the DMABz4 is blue-green ($\lambda_{max} = 500\ nm$). Photoluminescence quantum yields (PLQYs) of the deoxygenated SBABz4 and DMABz4 solutions are 55.7% and 83.8% respectively (Table S1). Altogether, spiro-linkage in the SBABz4 core provides blue shift of the emission, while preserving reasonably high PLQY.

Currently prevailing TADF design with core-acceptor allows PLQY improvement via an increase in the number of donor units or the donor conjugation length.[2,18,22,27] The resulting bulky donor inevitably causes a red shift of luminescence, because of less localized and/or less stabilized HOMO. Alternatively, the rigid spiro-linkage maintains a large dihedral angle between the acridine subunits in the SBABz4 core, thus breaking the overlap between donor parts and introducing the desired blue shift into the emission. At the same time, the doubled number of emissive units still allows to maintain high luminescence efficiency. Therefore, SBABz4 was chosen for further study and fabrication of pure blue OLEDs.

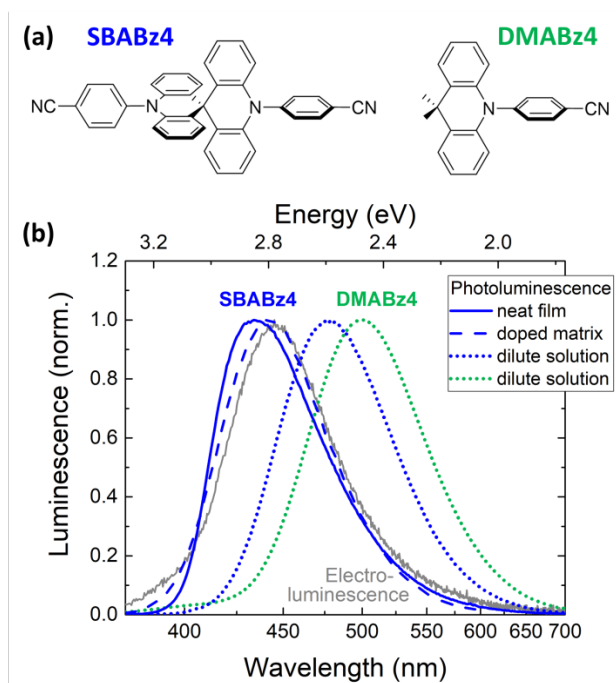

**Figure 1.** (a) Structural formulas of SBABz4 and DMABz4 emitters. (b) Normalized PL spectra of SBABz4 (blue) and DMABz4 (green). Solid, dashed, and dotted colored lines represent neat film, doped 5wt% SBABz4:DPEPO matrix, and dilute solutions in dichloromethane (DCM), respectively. Grey solid line shows electroluminescence from SBABz4-based OLED.

Figure 2 demonstrates transient PL and normalized PL spectra at different stages of decay for a solid 5wt% SBABz4:DPEPO film at room (298 K) and low (77 K) temperatures. Transient PL at both temperatures contains fast and slow components. The fast component, peaking at 435 nm, dominates within the first 100 ns of decay regardless of the temperature. We ascribe this component to the prompt fluorescence from the $^1$CT level. PL spectrum of the slow component at 77 K has noticeable red shift relative to the prompt one. Therefore, we assume that the slow component at low temperature is phosphorescence from lower-lying triplet. In contrast, the spectrum of the slow PL component at room temperature practically coincides with the prompt fluorescence. Thus, the delayed emission at room temperature originates from the same $^1$CT level as the prompt fluorescence. Therefore, we assign it to TADF, mediated by slow reverse intersystem crossing. The corresponding energy diagram is depicted in the Figure 2. Neat SBABz4 film qualitatively showed the same behavior (Figures S5-6).

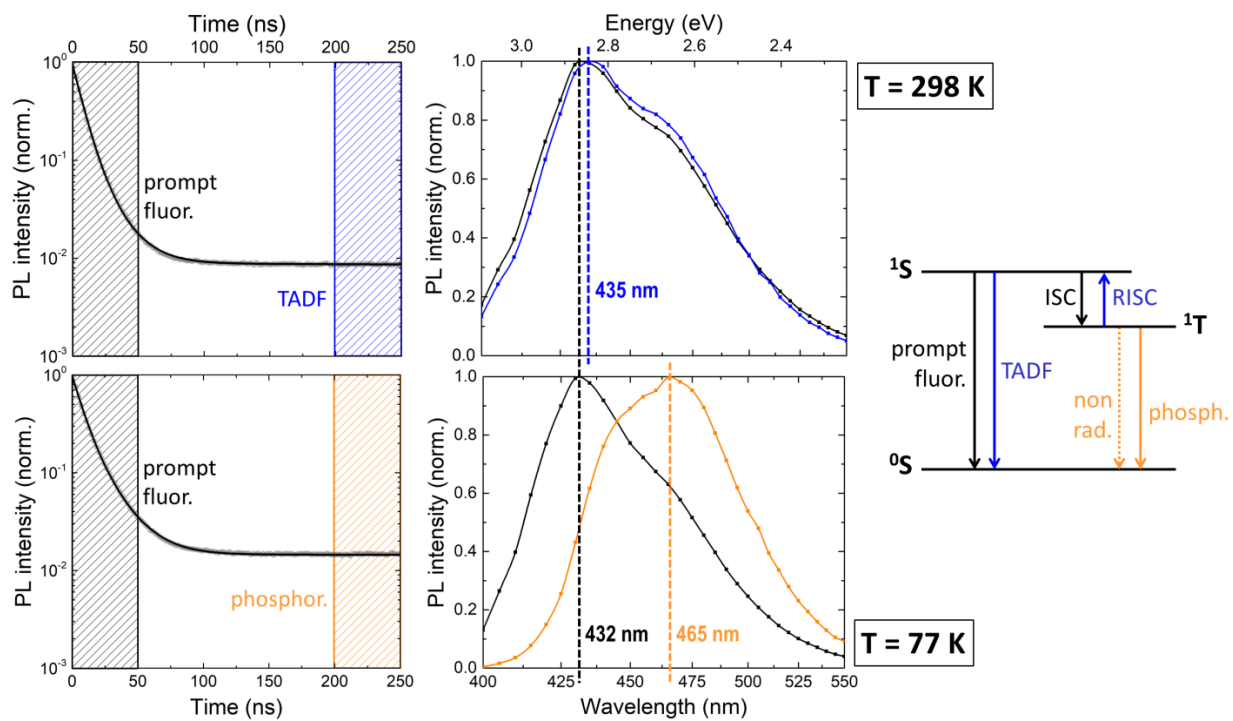

**Figure 2.** PL decay curves and the corresponding normalized PL spectra at different stages of decay in 5wt% SBABz4:DPEPO film at 298 K (upper plots) and 77 K (lower plots). Prompt components (black curves) were integrated within 0-50 ns, delayed (blue) and phosphorescence (orange) within 200-250 ns. Right: proposed energy diagram of the respective processes.

Figure 3 shows steady-state and time-resolved emission from a dilute solution of SBABz4 in DCM before and after oxygen exposure. Whereas the shape and position of PL spectrum remained unchanged under oxygen exposure, PLQY decreased sevenfold (from 55.7 to 8.2%, Figure 3a and Table S1). Note that the corresponding decay curves were recorded in the broad dynamic range of 5 orders of magnitude in time and 6 orders of magnitude in intensity. The transient PL contains prompt and delayed components (Figure 3b) with identical spectra (Figure 3b, inset). The fast component remains unaltered in the presence of oxygen. Therefore, we ascribe it to prompt fluorescence from $^1$CT. In contrast, the slow component is substantially quenched by oxygen. Therefore, we assign it to TADF from $^1$CT level populated via RISC. Figure 3c shows the same PL decay curve from the deoxygenated solution on a semi-logarithmic scale, where the straight lines indicate the purely monoexponential character of both decay processes. Characteristic time constants (16.5 ns for the prompt fluorescence and 19.6 μs for TADF), were extracted from the double-exponential fit, fully consistent with luminescence from simple three-level system with triplet up-conversion (Figure 3c).

Contribution of the prompt and delayed components in the total PLQY can be calculated numerically as the area under the decay curve or, in the simple case of monoexponential fit, expressed through the magnitudes and lifetimes of the components (SI, p.16):

$$\varphi_{pr} = \frac{A_{pr}\tau_{pr}}{A_{pr}\tau_{pr} + A_{TADF}\tau_{TADF}} \cdot PLQY, \qquad (1)$$

$$\varphi_{TADF} = \frac{A_{TADF}\tau_{TADF}}{A_{pr}\tau_{pr} + A_{TADF}\tau_{TADF}} \cdot PLQY; \quad (2)$$

Here $\varphi_{pr}$ and $\varphi_{TADF}$ are quantum efficiencies of the prompt fluorescence and TADF; $A_{pr}$ and $A_{TADF}$ are magnitudes of the decaying exponents; $\tau_{pr}$ and $\tau_{TADF}$ are respective lifetimes; $PLQY$ is the measured total photoluminescence quantum yield. According to the Equations (1-2) and fitting parameters, quantum efficiency of TADF in the SBABz4 ($\varphi_{TADF} = 0.420$) significantly exceeds that of prompt fluorescence ($\varphi_{pr} = 0.137$). Now, under the assumption that phosphorescence is not present at room temperature, quantum efficiencies of ISC, RISC, and non-radiative triplet decay can be expressed via $\varphi_{pr}$ and $\varphi_{TADF}$: $\varphi_{ISC} = 1 - \varphi_{pr}$, $\varphi_{RISC} = \varphi_{TADF}$, and $\varphi_{n.r.} = \varphi_{ISC} - \varphi_{TADF}$ (Figure 3d). With that, maximum internal electroluminescence efficiency of the emitter in OLED can be expressed as:[1]

$$\Phi_{EL,int} = \eta_S \varphi_{pr} + \eta_S \varphi_{ISC} \varphi_{RISC} + \eta_T \varphi_{RISC}; \quad (3)$$

Here $\eta_S$ and $\eta_T$ are portions of singlets and triplets produced via electrical injection (0.25 and 0.75, respectively). We obtained $\Phi_{EL,int} = 44\%$ for SBABz4, which results in the estimation of $EQE_{max} = 8.8\%$ in the devices with assumed 20% light outcoupling. The estimation exceeds the 5% limit for first generation OLEDs. A fabricated SBABz4-based device with non-optimized layer thicknesses and doping concentration showed $EQE_{max} = 6.8\%$ close to the estimation (Figure S9). The device physics will, however, be explicitly discussed elsewhere.

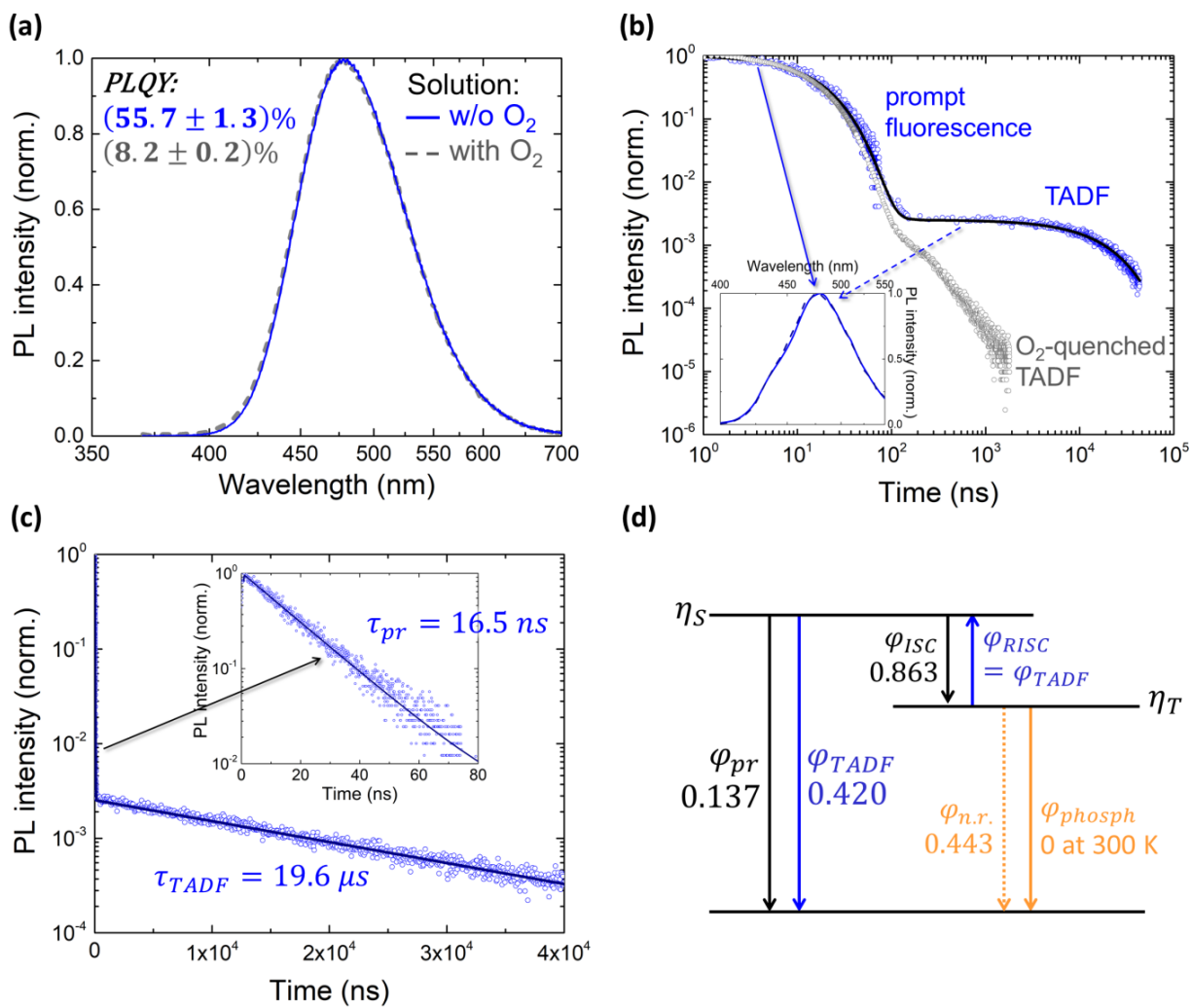

**Figure 3.** Steady-state and transient PL of dilute ($10^{-5}$ mol L$^{-1}$) oxygenated (grey) and deoxygenated (blue) solutions of SBABz4 in DCM at room temperature. (a) PL spectra normalized at the peak value. (b) Transient PL detected at 475 nm. The inset: normalized PL spectra recorded within 0-25 ns (solid) and 150-1500 ns (dashed) of the decay. (c) Transient PL of the deoxygenated solution on a semi-logarithmic scale. Inset: the prompt component. (d) Energy diagram depicting quantum efficiencies of the involved processes.

Figure 4 demonstrates PL decay curves of the neat SBABz4 film and 5wt% SBABz4:DPEPO film. Transient PL in both films clearly shows two distinctive components: prompt fluorescence and TADF. Both components deviate from the purely exponential decay behavior observed previously in the dilute solution (Figure 3c). We found out that a stretched exponent provided an excellent fit for both prompt and delayed components. Hence, the decay curves were fitted with a sum of two stretched exponents, and characteristic lifetimes were extracted as their average decay times (SI, p.17 and Table S2). These lifetimes are not directly comparable to the lifetime of single-exponential PL decay in the solution; therefore, we denoted them as $\langle \tau_{pr} \rangle$ and $\langle \tau_{TADF} \rangle$.

Notably, delayed fluorescence in the neat film decays two orders of magnitude faster than in the host matrix (0.13 $\mu$s and 9.8 $\mu$s, respectively). We assign the increased decay rate in the neat SBABz4 film to molecular aggregation with subsequent increase of non-radiative decay rate. For the same reason, PLQY of the doped matrix is higher than this of the neat film (40% and 27%, respectively; Figure 4 and Table S1).

The ratio between quantum efficiencies of the prompt and delayed components $\varphi_{pr}/\varphi_{TADF}$ in the films was calculated as the ratio of the areas under the corresponding curves (Table S2). Remarkably, prompt fluorescence in the neat film outperforms TADF ($\varphi_{pr}/\varphi_{TADF} \approx 5.83$), whereas both components equally contribute to the emission of the doped matrix ($\varphi_{pr}/\varphi_{TADF} \approx 1.04$). Now, using the measured values for PLQY, one can extract absolute quantum efficiencies of the involved processes for thin films (Figure S8). Indeed, TADF efficiency drops dramatically in the neat film ($\varphi_{TADF} \approx 0.040$) in comparison to the doped matrix ($\varphi_{TADF} \approx 0.196$). All in all, isolating and immobilizing the emitter molecules in the host matrix significantly improve TADF lifetime and efficiency, which is essential for OLED performance.

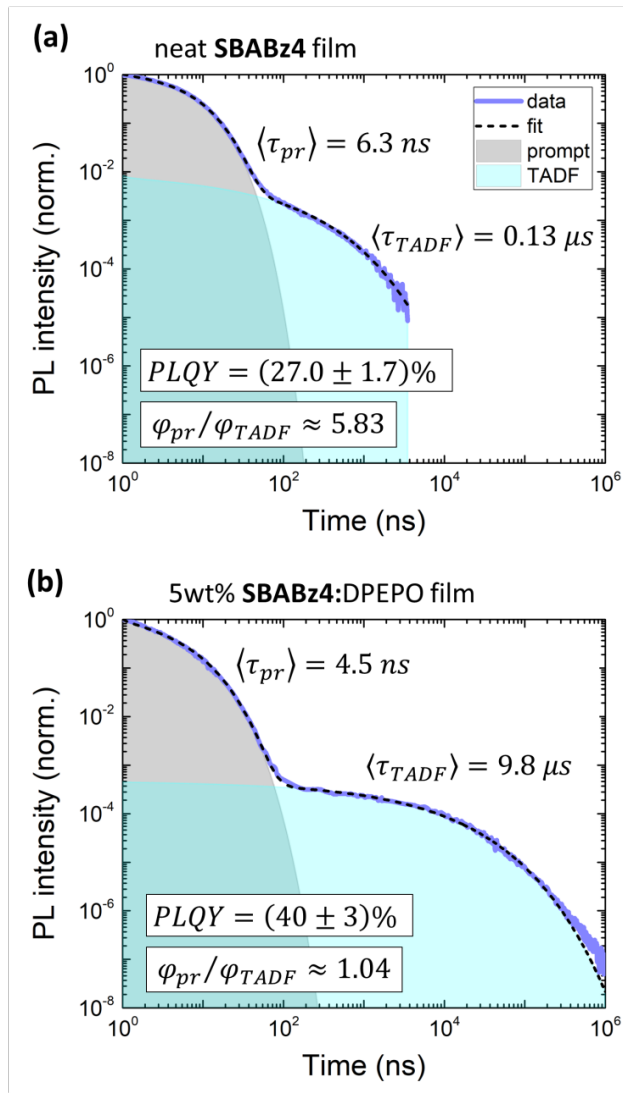

**Figure 4.** PL decay curves of the neat SBABz4 (a) and 5wt% SBABz4:DPEPO (b) films detected at 435 nm at room temperature. Fit (dashed) is the sum of two stretched decaying exponents.

We applied Equation (3) in order to obtain performance estimates for SBABz4-based OLEDs from photophysical constants of thin films (Table S3). Resulting maximum internal EL efficiency of 5wt% SBABz4:DPEPO film exceeds that of the neat film more than twice (0.237 and 0.095, respectively). Interestingly, the estimated EQE for the OLED based on the doped matrix ($EQE_{max} = 7.1\%$) is in excellent agreement with the corresponding experimental value ($EQE_{max} = 6.8\%$), as long as the light outcoupling of 30% is assumed (Table S3). Therefore, we propose that the light outcoupling of SBABz4-based OLEDs exceeds conventional 20%, as it was shown earlier for other stick-like emitters.[5,6,28] Although self-orientation of the SBABz4 emitter lies outside the scope of this work, the problem needs to be elucidated in future studies.

Temperature-dependent TRPL in combination with absolute PLQY measurements allows calculation of the energy splitting $\Delta E_{ST}$ for the singlet and triplet levels involved in TADF process (SI, p. 20-23; Figure S10, S11).[29] We obtained $\Delta E_{ST} \approx 70$ meV for 5wt% SBABz4:DPEPO film. This energy difference is sufficiently low ($\Delta E_{ST} \sim k_B T$) to provide efficient TADF. $\Delta E_{ST}$ can be further decreased via judicious variation of acceptor units. On the other hand, stronger acceptors were shown to cause red shift of emission in similar systems,[5] whereas the emission from SBABz4 is deep blue.

In conclusion, we designed and synthesized a novel compact deep-blue OLED emitter, SBABz4, containing spiro-biacridine and benzonitrile units, and its monomer counterpart DMABz4. We showed that the spiro-linkage in the double-donor core of SBABz4 renders its luminescence pure-blue relative to the single-donor DMABz4, while preserving efficient TADF. Therefore, core-donor provides desirable for applications color tuning in the deep blue region, as opposed to the commonly used TADF molecular design with core-acceptor. The SBABz4 emitter combines prompt fluorescence, phosphorescence at low temperature, and TADF at room temperature. The latter was independently proven by temperature-dependent transient PL measurements and oxygen-quenching of the delayed PL component. We estimated $EQE_{max} = 7.1\%$ in SBABz4-based OLEDs from the PL lifetimes and efficiencies, and obtained $EQE_{max} = 6.8\%$ in an operating test device. Transient PL was recorded from the solutions and solid films in the broad dynamic range covering up to 8 orders of magnitude in intensity and 6 orders of magnitude in time. The stretched exponent is shown to fit the transient PL in the films very well, whereas PL decay in dilute solution is found to be purely exponential. Immobilization and isolation of the emitter molecules in the DPEPO host matrix maintain the efficient TADF. Finally, we show that SBABz4 emitter doped in the host matrix demonstrates superior photophysical properties over the neat film.

EXPERIMENTAL METHODS

**Solution preparation**

Deoxygenated solutions of SBABz4 and DMABz4 in dichloromethane (DCM) with concentration $10^{-5}$ mol L$^{-1}$ were prepared in a glove box under nitrogen atmosphere.

Deoxygenated solutions of SBABz4 for UV-vis absorption and luminescence measurements in different solvents (Figure S4) were obtained by bubbling nitrogen and sealing with a JYoung valve. Spectroscopic grade solvents were used without further purification.

**Film preparation**

Films for optical measurements were prepared in a glove box with nitrogen atmosphere. Neat SBABz4 and SBABz4:DPEPO films were processed on a 0.5x0.5 sq.-inches glass substrates without modification. The substrates were sequentially cleaned with deionized water, acetone, and isopropanol for 10 min each in an ultrasonic bath and plasma-edged for 30 s before use. Solutions of SBABz4 (or SBABz4:DPEPO with the desired weight ratio) in DCM with concentration 5 mg mL$^{-1}$ were spin-coated at 3000 rpm during 1 min. After spin-coating, the substrates were annealed on a hot plate at 100°C for 10 minutes.

**PL measurements**

PL spectra and transients of solutions and films were recorded with calibrated Fluorescence Spectrometer FLS980 (Edinburgh Instruments), equipped with three excitation sources: continuous broad-spectrum xenon lamp Xe1, pulsed violet diode laser EPL-375 (wavelength 375.0 nm, pulse width 80 ps), and microsecond xenon flash lamp $\mu$F920. Laser repetition rate was tunable with discrete steps in the range from 2.5 kHz to 20 MHz. Spectral resolution of the system was 0.1 nm. Temporal resolution for time-resolved measurements varied from 2 to 50 ns, depending on the recorded time range.

All measurements for solid films at room and low temperatures were carried out in active vacuum ($10^{-5}$ mbar) in Janis ST-100 cryostat. Measurements at room temperature were performed beforehand for every sample. Emission recorded at different time scales was merged to improve temporal resolution for the prompt component.

PLQYs were measured with the calibrated integrating sphere F-M01 in FLS980 spectrometer. All measurements for solutions were conducted in a standard (10x10 mm) quartz cuvette with a tight screw cap. Solid films for PLQY measurements were sealed in a quartz cuvette under nitrogen atmosphere.

ASSOCIATED CONTENT

**Supporting Information Available:** synthetic details, thermal and electrochemical characterization, additional photophysical data, equations and parameters for TRPL fits, performance estimates for OLEDs, OLED layout, equations and experimental results for RISC activation energy, including Schemes S1-S3, Tables S1-S3, and Figures S1-S11 (PDF).

AUTHOR INFORMATION

**Notes**

The authors declare no competing financial interests.

ACKNOWLEDGMENT

L.G.K., A.S., and V.D. acknowledge the EU H2020 for funding through the grant SEPOMO (Marie Skłodowska-Curie grant agreement No 722651). S.W. acknowledges the funding from DFG FOR1809.